# A Framework for the High-Level Specification and Verification of Synchronous Digital Logic Systems[*]

Nick Mertin[†]

Department of Electrical and
Computer Engineering
Queen's University
Kingston, Ontario, Canada
17nfam@queensu.ca

K. Ritsuka

Department of Electrical and
Computer Engineering
Queen's University
Kingston, Ontario, Canada
ritsuka314@queensu.ca

Karen Rudie

Department of Electrical and
Computer Engineering
Queen's University
Kingston, Ontario, Canada
karen.rudie@queensu.ca

**Abstract**

A syntactic model is presented for the specification of finite-state synchronous digital logic systems with complex input/output interfaces, which control the flow of data between opaque computational elements, and for the composition of compatible systems to form closed-loop systems with no inputs or outputs. This model improves upon similar existing models with a novel approach to specifying input and output ports in a way which is uniform and symmetric. An automaton model is also presented for encoding arbitrary computational processes, and an algorithm is presented to generate an automaton representation of a closed-loop system. Using the automaton model, the problem of timing-agnostic verification of closed-loop systems against a desired behavioural specification—encoded as the similarity of closed-loop systems in terms of the set of computations performed—is shown to be decidable. The relationship between the models and real-world implementations of systems is discussed.

## 1 Introduction

Nearly every part of a modern digital electronic system, from CPU cores and digital signal processors to storage units and communication interface controllers, is a synchronous digital logic system, consisting of a combination of inputs, outputs and memory elements which are connected by combinational logic functions and driven by a clock signal. Designing these systems and the connections between them is a complicated task, which in practice involves a combination of cleverness on the part of the designer and software tools for electronic design automation (EDA).

Verifying the designs of digital electronics systems is an even more difficult task, for a combination of reasons: firstly, the designs themselves are immensely complicated, resulting in state-space explosion and other scaling effects when attempting naive verification methods or verification by simulation, and secondly, precisely encoding the range of acceptable behaviour in a form that can be reasoned about by the verification tool is itself a challenge. For example, consider the task of designing a CPU core which implements an instruction-set architecture. There are practically infinitely many possible valid designs, with slight differences between them: some designs include only one arithmetic and logic unit (ALU), while others have multiple, allowing for improved performance; some only execute instructions once it is known that they will need to be executed, while others perform reads, arithmetic operations, or other side-effect-free instructions speculatively in order to improve performance. While there may be additional, quantitative verification steps to assess the performance trade-offs, it would be ideal if all of these designs could be formally verified against a single, "timing-agnostic" specification of the behaviour required by the instruction-set architecture, which allows for these optimizations—but prohibits, for example, speculatively writing to memory, as that could change subsequent behaviour and would therefore be a logical error.

### 1.1 Related Work

Existing verification methods take a variety of approaches, depending on the intended level of abstraction. Various tools, such as ABC [8], exist for verifying sequential Boolean logic against a cycle-by-cycle specification of outputs as a function of present and past inputs. Other methods, particularly those built around term rewriting systems [1, 14], are more aimed at timing-agnostic verification; however, these works are very focused on the specifics of designing RISC processors and do not result in a generic model applicable to all digital logic design. Finally, tools such as DFiant [21] enable timing-agnostic synthesis of hardware, but do not provide for formal verification of arbitrary designs.

On the theoretical front, existing work presents a variety of formal models used to describe digital logic. Brown and Hutton [9] present a robust algebraic model for purely combinational digital logic, however there is no direct extension of the model to sequential circuits. Braibant [7] presents a model of circuits built in the Coq proof assistant, which does support sequential logic, but does not provide a model for specifications of desired behaviour beyond noting that one may define and attempt to prove arbitrary propositions about a circuit within Coq. Rudie and Wonham [22] present

---

a solution for a related problem (the verification of communication protocols), which omits the flow of data itself from the model and instead works only with asynchronous events and control decisions; the interpretation of these as relating to the flow of information is not part of the model.

Various other works model synchronous digital logic systems as coalgebras [25, 3, 4, 5, 2, 28, 24, 16, 29, 18], a concept arising from category theory which is effectively a generalization of finite state automata, regular expressions, term rewriting systems, and similar constructions; coalgebras have previously been used as a modelling formalism in related issues such as the semantics of programming languages [26]. However, while there has been significant work put into modelling and specifying a wide range of systems using coalgebras, none of these approaches consider the need for timing-agnostic verification or the requisite behavioural specifications.

### 1.2 Contributions

The contributions of this paper are as follows:

- A typed syntax for coalgebras which describe both arbitrary synchronous digital logic systems and desired behaviours thereof. This model is largely based on the work of Bonsangue *et al.* [3, 4, 5], but is augmented to support computation with arbitrary opaque types and functions and enhanced with a composition operator.
- A formulation of automata which describe computational processes using the opaque types and functions described above. This model is used as an interpretation of the semantics of systems in the syntactic model.
- Compatible definitions for similarity on both the syntactic systems and automata which formally encode the idea of a system conforming to a specification of desired behaviour. It is shown, using automata, that the similarity of arbitrary systems is decidable; in other words, given a design and a behavioural specification, it is always possible to determine whether the design conforms to the specification.
- Discussions throughout relating the formal models to real-world implementations.

Section 2 provides a brief theoretical background on category theory and coalgebraic models of dynamic systems. Section 3 describes the base system of opaque types and functions and shows how a model similar to the polynomial functors in [4] is useful to model the interfaces of synchronous digital logic systems. Section 4 describes the typed syntax used to specify systems and behaviours. Section 5 describes the automata used to model computational processes and provides the decidability proof for similarity. Section 6 explains how the results may be applied to systems which interact with the outside world. Finally, Section 7 discusses potential directions of future work.

## 2 Background

This section provides a sufficient background on category theory to understand the models and proofs presented in this paper, beginning with definitions for functors and coalgebras, and then discusses some of the related works in more technical detail.

**Definition 2.1** (Functor). In category theory, a *functor* (or more precisely, an endofunctor on sets) consists of [6, p. 5]:

1. a mapping from sets to sets, $F : \mathbf{Set} \Longrightarrow \mathbf{Set}$, where the image of a set $X$ is conventionally written $FX$; and
2. for any sets $A$ and $B$ and function $f : A \longrightarrow B$, a function $Ff : FA \longrightarrow FB$; this is often called $f$ "lifted through" the functor $F$.

**Example 2.2.** The set of all strings over an alphabet $\Sigma$ is commonly denoted $\Sigma^*$. This $-^*$ operator is in fact a functor: for any set $X$, its image $X^*$ is also a set, and for any function $f : X \longrightarrow Y$, the function $f^* : X^* \longrightarrow Y^*$ maps $f$ element-wise over each symbol in the string.

Often, the mapping of functions through a functor is left implicit, and the functor is given solely by its mapping of sets.

**Definition 2.3** (Coalgebra). Given a functor $F$, a *coalgebra* on $F$, or $F$-*coalgebra*, is a pair $(S, \eta)$ consisting of a set $S$ and a function $\eta : S \longrightarrow FS$. The set $S$ is the *state space* or *carrier set* of the coalgebra, and the function $\eta$ describes the dynamics of the system [25, 5].

**Example 2.4.** Let $\Sigma$ be an alphabet, $\mathrm{Re}_\Sigma$ be the set of regular expressions over $\Sigma$, and $\mathbb{B} = \{0, 1\}$ be the set of Boolean values. Suppose $F$ is the functor which maps a set $X$ to $X^\Sigma \times \mathbb{B}$, i.e., to pairs of a function from symbols in the alphabet to $X$ and a Boolean value. Suppose that $\eta$ is a function which maps a regular expression $r$ to the pair $(\lambda \sigma. D_\sigma(r), E(r))$, where $D_\sigma(r)$ is the derivative of $r$ with respect to $\sigma$, per Brzozowski [10], and $E(r)$ is the Boolean-valued function which indicates whether $r$ accepts the empty string. Then, $(\mathrm{Re}_\Sigma, \eta)$ is an $F$-coalgebra, which describes the complete dynamics of regular expressions as a description of (deterministic) finite automata [16].

Note that the coalgebra described in Example 2.4 does not describe the dynamics of just one regular expression, but of all of them; its state space is not the set of states in a single automaton, but rather in all deterministic finite automata. Despite the carrier set $\mathrm{Re}_\Sigma$ being infinite, it just so happens that, as it describes only regular languages, its elements can be partitioned (per Nerode [20]) into equivalence classes of expressions denoting the same language such that, from any state, only a finite subset of equivalence classes ("intrinsic states") are reachable.

Much of the relevant work cited focuses on creating and reasoning about analogous systems of regular expressions and coalgebras on broad classes of functors [4, 5, 2, 28]. These efforts are successful at that task, however the corresponding Myhill-Nerode equivalence relation, which allows for all systems to be implemented using finite state spaces, is—as with traditional regular expressions—a complex combination of commutativity, associativity, and similar syntactic properties, which makes mapping a specification to hardware difficult and sometimes unintuitive.

Moreover, as with traditional regular expressions, there is no concept in these models of storing or passing an opaque value; rather, all inputs must be explicitly discriminated on, and all outputs must be explicitly given. This significantly complicates modelling data flow, such as in communication systems, requiring one to either face state-space explosion or omit data flow from the model (as in [22]).

## 3 Framework

### 3.1 Types and Data

As noted earlier, our model describes computation in terms of opaque types and functions. These symbols may have any interpretation; for example, the types could be integer types of various sizes, and the functions could be conversions and arithmetic operations. However, the interpretations are irrelevant to the model (hence "opaque"); the opaque functions constitute the atomic elements of computation which may be performed, and the type system constrains how data may be passed between those functions.

Let $\mathbb{T}_\bullet$ be the set of opaque types (where the subscript $\bullet$ indicates opacity). We extend this to a richer set of types $\mathbb{T}$, called composite types, by introducing the singleton type $1 = \{*\}$, as in [4], and allowing types to be arbitrarily combined by finite product (with infix $\times$) and finite coproduct (with infix $+$). The notation for composite types is given by the BNF:

$$\langle\text{type}\rangle \quad ::= \quad \text{an opaque type} \mid 1 \mid$$
$$\langle\text{type}\rangle\times\langle\text{type}\rangle \mid \langle\text{type}\rangle+\langle\text{type}\rangle$$

Opaque functions are maps from one composite type (the domain) to another composite type (the codomain). Let $\mathbb{F}_\bullet$ be the set of opaque functions; we denote the domain and codomain of opaque functions using $\text{dom} : \mathbb{F}_\bullet \longrightarrow \mathbb{T}$ and $\text{cod} : \mathbb{F}_\bullet \longrightarrow \mathbb{T}$, respectively; i.e., for any $f \in \mathbb{F}_\bullet$, we have $\text{dom}\, f, \text{cod}\, f \in \mathbb{T}$. Together, these comprise the *signature* of the function, written in the form $f : \text{dom}\, f \longrightarrow \text{cod}\, f$.

**Example 3.1.** Suppose $\text{int} \in \mathbb{T}_\bullet$ is a type corresponding to integers of a certain size. A function $\text{add} \in \mathbb{F}_\bullet$ representing binary addition of these integers would likely have the signature $\text{int} \times \text{int} \longrightarrow \text{int}$ (i.e., $\text{dom}\,\text{add} = \text{int} \times \text{int}$ and $\text{cod}\,\text{add} = \text{int}$).

This type system is very similar to that which might be found in a bare-bones functional programming language, albeit without any capability of modelling data structures of unbounded size (e.g., inductive types).

To facilitate the automaton model presented in Section 5, we also provide the following definition:

**Definition 3.2** (Sum-of-Products Form and Variants). Any type in our type system may be converted to a *sum-of-products form*, where a coproduct is never a component of a product, which is isomorphic to the original type. This is done by recursively distributing products over coproducts (sums), exactly as in expanding a polynomial or computing the disjunctive normal form of a propositional expression.

Throughout, we will refer to the product subterms of the sum-of-products form of a type as the *variants* of that type. The set of all variants of a type $T$ is denoted $\text{var}\, T$, and for any variant $v \in \text{var}\, T$, the list of components in that variant is denoted $\text{typ}\, v \in \mathbb{T}_\bullet^*$.

*Remark.* Different variants of a type may have identical type lists. To ensure that variants are distinguishable, they are identified by their (zero-based) index rather than by the types they contain.

**Example 3.3.** Let foo be an opaque type and $T$ be the composite type $\text{foo} \times (1 + 1)$. The sum-of-products form of $T$ is $\text{foo} \times 1 + \text{foo} \times 1$; it has two variants, both of which have a component list consisting solely of foo.

Though both variants have the same type lists, that does not make them redundant; distinguishing between the variants is itself information which is carried by values of the type, as is clearly the case with $1 + 1$, which is effectively a "transparent" encoding of a Boolean type.

### 3.2 Interface Shapes

The interfaces of synchronous digital logic systems are formalized as functors on the type system, such that a system is a coalgebra on the functor representing its interface, as in [4]. This functor is called the *shape* of the system.

The class of polynomial functors we consider is similar to that which is presented in [4] and is given by the BNF:

$$\langle\text{ftor}\rangle \quad ::= \quad \text{Id} \mid \langle\text{ftor}\rangle\times\langle\text{ftor}\rangle \mid \langle\text{ftor}\rangle+\langle\text{ftor}\rangle \mid$$
$$\langle\text{ftor}\rangle^{\langle\text{type}\rangle} \mid \langle\text{ftor}\rangle_{\langle\text{type}\rangle}$$

We denote the set of these functors by $\mathcal{F}$. Here, Id is the identity functor (which, alone, represents the shape of a system with no inputs or outputs); infix $\times$ and $+$ represent products (i.e., pairs) and coproducts (i.e., discriminable disjoint unions, not the extended definition used in [4]), respectively; and superscript and subscript represent input and output ports of the given type, respectively. The order in which these constructors are applied to create a functor describes the (possible) dependency between the values of the ports they represent; a more outer port is described as being *before* a more inner port, as it is encountered earlier when recursing through the structure of a shape, and therefore its value must be known earlier in a cycle and cannot depend on the value (in that same cycle) of a port which is after it.

**Example 3.4.** Two well-known classes of synchronous digital logic systems are Moore machines and Mealy machines. Both have one input port and one output port, but whereas the output of a Moore machine depends only on its current internal state, the output of a Mealy machine can depend on both the internal state and the current inputs. Suppose $T, U \in \mathbb{T}$ are types; the shape of a Moore machine with an input port of type $T$ and an output port of type $U$ would be represented by the expression $\left(\text{Id}^T\right)_U$, while the shape of an equivalent Mealy machine would be represented by the expression $(\text{Id}_U)^T$.

In our model, both products and coproducts represent a system which has the ability to be one of two different shapes, given by the two operands; the difference between them is that for a product functor, the choice between the possible shapes is an input, whereas for a coproduct functor, it is an output. These semantics provide us with a rudimentary form of dependent typing, which enable the precise representation of complex interfaces, such as those found in transceivers for communication busses (see Example 6.1), as well as the interconnection of multiple components.

We also define the *complement* of a functor:

**Definition 3.5** (Complement of a Functor). For a polynomial functor $F$, its *complement*, written $\overline{F}$, is defined inductively on the syntactic structure of the functor as follows:

$$\overline{\text{Id}} = \text{Id}$$
$$\overline{F_1 \times F_2} = \overline{F_1} + \overline{F_2}$$
$$\overline{F_1 + F_2} = \overline{F_1} \times \overline{F_2}$$
$$\overline{F^T} = \overline{F}_T$$
$$\overline{F_T} = \overline{F}^T$$

Intuitively, the complement is the functor with all inputs replaced by equivalent outputs, and vice versa. In our model, this describes the relationship between the shapes of compatible systems to be composed together; the semantics of this are discussed in Section 4.3.

*Remark.* Unlike [4], our formulation of polynomial functors does not include constant functors as a means of encoding outputs. The reason for this is twofold. First, encoding output ports and input ports in an analogous form allows for the definitions of the complement—and related concepts in Section 4—to be straightforward and for the duality between inputs and outputs in the model to be clear. Second, an implication of including constant functors is that it becomes possible to specify coalgebras which could "terminate"; there is no clear analogue for termination in the concept of synchronous digital logic, and it would only complicate the semantics of the model.

### 3.3 Relation to Physical Hardware

The type system and interface shape models described in this section are intended to represent physical hardware concepts. As such, the opaque types generally represent information which has a binary representation which may be used on a bus or in a register, and opaque functions generally represent known combinational circuits. Potential exceptions to types and functions being representable in this way are discussed in Section 6.

Products and coproducts of representable types may be represented in a variety of ways, analogous to the in-memory representation of record and enumeration types in programming languages. Often, this means that a product type is represented by separate bits for each component, while a coproduct type is represented by a single-bit discriminator and a set of data bits which are shared by the variants.

Interface shapes are representable if and only if all types which appear in the interface specification are representable. The representation of the interface can be defined, similarly to that of types, in terms of the structure of the shape specification. The identity functor has no inputs or outputs, and therefore its representation is an "empty" interface, with no input or output lines. Input and output ports add input or output lines for the given data type to the representation of the interface.

**Example 3.6.** As the Moore and Mealy machine shapes $\left(\text{Id}^T\right)_U$ and $(\text{Id}_U)^T$ discussed in Example 3.4 have equivalent input and output types, under the implementation scheme proposed above, they would have isomorphic physical interfaces (i.e., the only difference would be the labelling of data lines). This aligns with classical understanding of these classes of machines: the difference between them is not in the physical interface, but rather in what is permitted internally and externally with respect to connections to that interface.

As both product and coproduct functors represent an interface which can take one of two possible shapes, they may both be implemented by a single-bit discriminator line and a set of data lines which are shared by the two variants. The direction of the discriminator line is as an input for products and as an output for coproducts. This demonstrates how products and coproducts are the primitive building blocks for half-duplex data busses; in general, given types $T, U \in \mathbb{T}$, a bus interface which may either transmit a value of type $T$ or receive a value of type $U$ each cycle has shape $\text{Id}^T \times \text{Id}_U$ if it has control over whether it is transmitting or receiving and $\text{Id}^T + \text{Id}_U$ if the bus has control over the direction.

*Remark.* There are many possible physical representations for any type or interface; an application of our model to real hardware must include a mapping to the desired representation. One could consider this mapping to be the analogue of a technology mapping algorithm in traditional EDA, or of a calling convention in high-level programming languages such as C.

## 4 Specification of Synchronous Digital Logic

### 4.1 Combinational Logic

While many models of combinational logic, such as [9], are built around the manipulation of bits, that is not the goal of our model. As we instead wish to model at a more abstract level of application of arbitrary opaque functions (assumed to be as primitive as the basic Boolean operations on bits) to values of opaque types, we instead present a model which is more akin to the simply-typed lambda calculus of Church [11], using an algebra of typed expressions to inductively build up functions from primitive operators and combinators.

The defining property of functions which can be implemented as combinational logic is that there is a bounded amount of computation performed, as all hardware involved is used at most once in the computation. When creating a model for specifying these functions, this property is equivalent to ensuring that the model does not permit the encoding of any form of looping or recursion; therefore, each syntactic element of the specification corresponds to a physical element which is used at most once in any evaluation of the function.

We use a rich algebra of expressions for defining functions, the syntax of which is given by the BNF:

$$\langle\text{func}\rangle ::= \text{an opaque function} \mid \langle\text{func}\rangle \circ \langle\text{func}\rangle \mid \\ \langle\langle\text{func}\rangle, \langle\text{func}\rangle\rangle \mid [\langle\text{func}\rangle, \langle\text{func}\rangle] \mid \\ \text{id} \mid \theta \mid \pi_1 \mid \pi_2 \mid \kappa_1 \mid \kappa_2 \mid \delta_1$$

These expressions are typed according to the rules given in Figure 1; the context $\Gamma$ is understood to contain the typing rules opaque functions. The combinators $- \circ -$, $\langle -, - \rangle$ and $[-, -]$ are, respectively, function composition, constructing a product, and discriminating on a coproduct (i.e., the usual meanings in category theory). The operators $\text{id}$, $\theta$, $\pi_1$, $\pi_2$, $\kappa_1$, $\kappa_2$ and $\delta_1$ are, respectively, the identity function, the map to the singleton type 1, the left projection of a product, the right projection of a product, the left constructor of a coproduct, the right constructor of a coproduct, and the left distributive law of products over coproducts. The duality of products and coproducts can be seen in the identities $\pi_1 \circ \langle f, g \rangle = f = [f, g] \circ \kappa_1$ and $\pi_2 \circ \langle f, g \rangle = g = [f, g] \circ \kappa_2$.

$$\frac{\Gamma \vdash f : T \longrightarrow U \quad \Gamma \vdash g : U \longrightarrow V}{\Gamma \vdash g \circ f : T \longrightarrow V}$$

$$\frac{\Gamma \vdash f : T \longrightarrow U \quad \Gamma \vdash g : T \longrightarrow V}{\Gamma \vdash \langle f, g \rangle : T \longrightarrow U \times V}$$

$$\frac{\Gamma \vdash f : T \longrightarrow U \quad \Gamma \vdash g : T \longrightarrow V}{\Gamma \vdash [f, g] : T + U \longrightarrow V}$$

$$\overline{\Gamma \vdash \text{id} : T \longrightarrow T} \qquad \overline{\Gamma \vdash \theta : T \longrightarrow 1}$$

$$\overline{\Gamma \vdash \pi_1 : T \times U \longrightarrow T} \qquad \overline{\Gamma \vdash \pi_2 : T \times U \longrightarrow U}$$

$$\overline{\Gamma \vdash \kappa_1 : T \longrightarrow T + U} \qquad \overline{\Gamma \vdash \kappa_2 : U \longrightarrow T + U}$$

$$\overline{\Gamma \vdash \delta_1 : T \times (U + V) \longrightarrow (T \times U) + (T \times V)}$$

**Figure 1.** Type inference rules for expressions defining combinational logic functions. All rules are implicitly universally quantified over the context $\Gamma$, the expressions $f$ and $g$, and the types $T$, $U$ and $V$.

As a convenience, we define the following additional combinators and operators:

$$f_1 \times f_2 = \langle f_1 \circ \pi_1, f_2 \circ \pi_2 \rangle$$
$$f_1 + f_2 = [\kappa_1 \circ f_1, \kappa_2 \circ f_2]$$
$$\sigma_\pi = \langle \pi_2, \pi_1 \rangle$$
$$\sigma_\kappa = [\kappa_2, \kappa_1]$$
$$\delta_2 = [\sigma_\pi, \sigma_\pi] \circ \delta_1 \circ \sigma_\pi$$

Intuitively, infix $\times$ and $+$ map a separate function over either component of a product or coproduct, respectively; $\sigma_\pi$ and $\sigma_\kappa$ swap the order of a product or coproduct, respectively; and $\delta_2$ is the right distributive law of products over coproducts. As these combinators and operators can be defined in terms of the primitive ones, doing so simplifies the expression model; they could be considered a "standard library" for building combinational expressions.

In a similar fashion to interface shapes, we can extend the physical representation presented in Section 3.3 to combinational expressions. The potential physical representations of the combinators and operators are fairly straightforward: function composition connects the inputs of one combinational circuit to the outputs of another; constructing a product places two combinational circuits in parallel, with the inputs connected, and the outputs separate; discriminating on a coproduct places two combinational circuits in parallel, with the inputs connected to the variants of the coproduct, and the outputs multiplexed based on the coproduct variant; the identity operator connects its inputs directly to its outputs; the singleton map does not use its inputs and has no outputs; the product projections connect a subset of their inputs directly to their outputs; coproduct constructors add a constant bit for the coproduct discriminator; and the distributive law simply rearranges bits such that the discriminator of the inner coproduct now applies to the outer coproduct.

### 4.2 Synchronous Logic

As with the interface shapes, our model of specifications for synchronous digital systems is inspired by that of [4]. Our model presents a somewhat simpler syntax and typing for expressions defining systems (which we will call *system expressions*, to differentiate from the combinational expressions introduced in Section 4.1). The syntax is given by the nonterminal $\langle\text{sys}\rangle$ in the BNF:

$$\langle\text{sysf}\rangle ::= \lceil\langle\text{func}\rangle\rceil \circ \omega \ | \ \alpha \circ \langle\text{sysf}\rangle \ |$$
$$\lceil\langle\text{func}\rangle\rceil \circ \beta \circ \langle\text{sysf}\rangle \ |$$
$$\lceil\langle\text{func}\rangle\rceil \circ (\langle\text{sysf}\rangle \otimes \langle\text{sysf}\rangle) \ |$$
$$\lceil\langle\text{func}\rangle\rceil \circ (\langle\text{sysf}\rangle \oplus \langle\text{sysf}\rangle)$$
$$\langle\text{sys}\rangle ::= \lceil\langle\text{func}\rangle\rceil \left(\mu(\langle\text{sysf}\rangle)\right)$$

Typing these expressions is more complex, as it requires the notion of a type of a system. Similarly to [4], we define $\triangleleft$ as the smallest reflexive and transitive relation on functors such that, for all $F_1, F_2 \in \mathcal{F}$ and $T \in \mathbb{T}$:

$$F_1 \triangleleft F_1 \times F_2 \qquad F_2 \triangleleft F_1 \times F_2 \qquad F_1 \triangleleft (F_1)^T$$
$$F_1 \triangleleft F_1 + F_2 \qquad F_2 \triangleleft F_1 + F_2 \qquad F_1 \triangleleft (F_1)_T$$

*Remark.* Unlike in [4], under our model, $\forall F \in \mathcal{F}. \ \text{Id} \triangleleft F$, due to the absence of constant functors. This is related to the fact that the class of system interfaces given in Section 3.2 only describe non-terminating machines.

We also define a syntax for system types as the BNF:

$$\langle\text{syst}\rangle ::= \lceil\langle\text{ftor}\rangle \triangleleft \langle\text{ftor}\rangle\rceil^{\langle\text{type}\rangle}$$

We require the instance of $\triangleleft$ in the notation to hold for the type to be valid. The superscript $\langle\text{type}\rangle$ is the type of the information held in the system (i.e., a parameter type).

We can now produce the typing rules for system expressions, which are given in Figure 2. The "ceiling" notation $\lceil-\rceil$ denotes "lifting" a function between data types to a function between system types; the combinator $-\circ-$ is function composition as usual; the operators $\alpha$ and $\beta$ represent accepting an input and producing and output, respectively; $\mu$ is the least fixed point combinator, representing the repetition of the combinational computation each clock cycle; $\omega$ maps a behaviour for the next iteration into a result of the current iteration, representing the memory element in the system; the combinator $-\otimes-$ allows for the construction of product results; and the combinator $-\oplus-$ allows for construction of coproduct results, by discriminating on a coproduct value.

*Remark.* When the function lifted by a ceiling notation is the identity function (i.e., when a term $\lceil\text{id}\rceil$ appears in a system expression), it is sometimes omitted for brevity.

*Remark.* Note that, for $f : T \longrightarrow U$, we lift $f$ to $\lceil f\rceil : [F \triangleleft G]^U \longrightarrow [F \triangleleft G]^T$; the swapping of $T$ and $U$ is intentional, as the type in the superscript denotes an "input" type, and therefore by attaching $f$ at the parameter input, we convert a system which takes $U$ as its parameter type to one which takes $T$. In category-theoretic terms, this describes a *contravariant* functor.

The form of the $\langle\text{sys}\rangle$ syntax applies a lifted computational function to a fixed point of a system specification function. The syntactic restrictions constrain the system specification function such that, regardless of the path taken in $\otimes$ and $\oplus$ combinators, it always describes exactly one "cycle" of the system (i.e., a single step in the coalgebra, or a single assignment of values to all the inputs and outputs and evaluation of combinational logic). In general, we call this function the *loop function* of the system, as the fixpoint causes it to be repeated forever; it is what may be interpreted as the function $\eta$ which characterizes the coalgebra. The parameter of its domain and codomain system types (i.e., the superscript $\langle\text{type}\rangle$ in the syntax for $\langle\text{syst}\rangle$) is called the *loop state type* of the system. Meanwhile, the lifted computational function which appears in the syntax for $\langle\text{sys}\rangle$ defines how the initial loop state is constructed from the parameter of the overall system; it is called the *initialization function*, and its input type—which is also the parameter in the overall system type—is called the *initial parameter type*.

**Example 4.1.** Suppose $T \in \mathbb{T}_\bullet$ is an opaque type and $\text{init} \in \mathbb{F}_\bullet$ is an opaque function with signature $1 \longrightarrow T$. Per Example 3.4, the shape of a Moore machine with $T$ as both the input and output type is $\left(\text{Id}^T\right)_T$. The expression for a system of this shape which implements a unit delay (i.e., its output each cycle is equal to its input in the previous cycle) with an initial value of $\text{init}(*)$ is:

$$\lceil\text{init}\rceil\left(\mu\left(\lceil\langle\theta,\text{id}\rangle\rceil \circ \beta \circ \alpha \circ \lceil\pi_2\rceil \circ \omega\right)\right)$$

This expression has type $\left[\left(\text{Id}^T\right)_T \triangleleft \left(\text{Id}^T\right)_T\right]^1$. The initial parameter type is the singleton type 1 since that is the domain of $\text{init}$, meaning that there is no additional initial state to be passed into the system, while the loop state type is $T$, as that is the type of the data which is stored between cycles.

**Definition 4.2** (Closed-Loop System). A *closed-loop system* is a system with the interface shape $\text{Id}$. Such a system represents a sequence of computational steps (i.e., applications of opaque functions to values) which is, in general, infinite, without any inputs or outputs.

$$\frac{\Gamma \vdash g : [F_1 \triangleleft G]^U \longrightarrow [F_2 \triangleleft G]^T \quad \Gamma \vdash x : [F_1 \triangleleft G]^U}{\Gamma \vdash gx : [F_2 \triangleleft G]^T} \qquad \frac{\Gamma \vdash g : [F_1 \triangleleft G]^V \longrightarrow [F_2 \triangleleft G]^U \quad \Gamma \vdash h : [F_2 \triangleleft G]^U \longrightarrow [F_3 \triangleleft G]^T}{\Gamma \vdash h \circ g : [F_1 \triangleleft G]^V \longrightarrow [F_3 \triangleleft G]^T}$$

$$\frac{\Gamma \vdash f : T \longrightarrow U}{\Gamma \vdash \lceil f \rceil : [F \triangleleft G]^U \longrightarrow [F \triangleleft G]^T} \qquad \overline{\Gamma \vdash \omega : [G \triangleleft G]^T \longrightarrow [\mathtt{Id} \triangleleft G]^T} \qquad \frac{\Gamma \vdash g : [G \triangleleft G]^T \longrightarrow [G \triangleleft G]^T}{\Gamma \vdash \mu g : [G \triangleleft G]^T}$$

$$\overline{\Gamma \vdash \alpha : [F \triangleleft G]^{T \times U} \longrightarrow [F^U \triangleleft G]^T} \qquad \overline{\Gamma \vdash \beta : [F \triangleleft G]^T \longrightarrow [F_U \triangleleft G]^{T \times U}}$$

$$\frac{\Gamma \vdash f : [F \triangleleft G]^U \longrightarrow [F_1 \triangleleft G]^T \quad \Gamma \vdash g : [F \triangleleft G]^U \longrightarrow [F_2 \triangleleft G]^T}{\Gamma \vdash f \otimes g : [F \triangleleft G]^U \longrightarrow [F_1 \times F_2 \triangleleft G]^T}$$

$$\frac{\Gamma \vdash f : [F \triangleleft G]^U \longrightarrow [F_1 \triangleleft G]^{T_1} \quad \Gamma \vdash g : [F \triangleleft G]^U \longrightarrow [F_2 \triangleleft G]^{T_2}}{\Gamma \vdash f \oplus g : [F \triangleleft G]^U \longrightarrow [F_1 + F_2 \triangleleft G]^{T_1 + T_2}}$$

**Figure 2.** Type inference rules for system expressions. All rules are implicitly universally quantified.

The construction of closed-loop systems from plants and controllers is discussed in Section 4.3. Since closed-loop systems are coalgebras on the identity functor, the usual concept of coalgebra simulations and bisimulations [17, 25] is trivial and not useful for comparing systems. Therefore, we produce the following definition, which is different from the usual definition in coalgebra and automata theory:

**Definition 4.3** (Similarity of Closed-Loop Systems). For any closed-loop system $S : [\mathtt{Id} \triangleleft \mathtt{Id}]^T$ and value $x_0 : T$, let $\Diamond(S, x_0)$ be the set of all pairs $(f, x)$, where $f \in \mathbb{F}_\bullet$ and $x : \mathrm{dom}\, f$, such that $S$ will eventually apply the function $f$ to the value $x$. Let $S_1$ and $S_2$ be closed-loop systems both having the initial parameter type $T$. Then, $S_1$ is *similar* to $S_2$, written by infix $\lesssim$, if and only if $\forall x_0 : T . \Diamond(S_1, x_0) \subseteq \Diamond(S_2, x_0)$, i.e., every computation performed by the first system is also performed by the second.

The result of this formulation of system specifications is a coalgebraic model of synchronous digital logic akin to that of [5], which is amenable to physical representations and able to directly model data flow applications.

### 4.3 Closed-Loop Composition of Systems

**Definition 4.4** (Closed-Loop Composition). Let $F \in \mathcal{F}$ be a polynomial functor and $T \in \mathbb{T}$ be a type. Suppose $S_1 : [F \triangleleft F]^T$ and $S_2 : [\overline{F} \triangleleft \overline{F}]^T$ are arbitrary systems of the given types. The *closed-loop composition* of $S_1$ and $S_2$, written $S_1 \circledast S_2$, has type $[\mathtt{Id} \triangleleft \mathtt{Id}]^T$ and is computed from the system expressions as shown in Figure 3.

The closed-loop composition describes the computational machine that is created when corresponding inputs and outputs of two compatible systems are connected. The compatibility is enforced by the requirement that the shapes of the two machines are complements of each other.

Based on the closed-loop composition, the following two problems are proposed:

**Problem 1** (Control Problem). Let $S$ be a closed-loop system, called the *specification*, and $P$ be a system of arbitrary shape, called the *plant*. Find a system $C$, called the *controller*, whose specification does not contain any opaque functions, such that $S \lesssim_\Diamond P \circledast C$.

**Problem 2** (Verification Problem). Given an instance of the control problem and a controller $C$, determine whether $C$ is a solution to the problem instance.

The control problem is not addressed in this paper. A solution to the verification problem is given in Theorem 5.16; the formulation of automata (detailed in Section 5) which implement closed-loop systems is critical to arriving at this solution.

## 5 Automata for Synchronous Digital Logic

This section describes a type of automata which we use to model the computational behaviour of closed-loop systems. The states in these automata have variables, whose types are opaque types, and the transitions describe how destination states are chosen based on source states and how the values of the variables in the destination state are computed from those in the source state. This model and the results produced from it demonstrate the power of the system specifications model given in Section 4.

The automata are formally defined as follows:

**Definition 5.1** (Automaton). An *automaton*, in the context of this model, is a 5-tuple $(Q, \rho, T_0, s_0, \tau)$, where:
- $Q$ is a finite set of states;
- $\rho : Q \longrightarrow \mathbb{T}_\bullet^*$ gives a finite list of types for the variables in each state;
- $T_0 \in \mathbb{T}$ is the initial parameter type;

$$\lceil f_1 \rceil \big(\mu(g_1)\big) \circledast \lceil f_2 \rceil \big(\mu(g_2)\big) = \lceil \langle f_1, f_2 \rangle \rceil \Big(\mu\big(g_1 \circledast_f g_2\big)\Big)$$

$$\Big(\lceil f_1 \rceil \circ \omega\Big) \circledast_f \Big(\lceil f_2 \rceil \circ \omega\Big) = f_1 \times f_2$$

$$(\alpha \circ g_1) \circledast_f \Big(\lceil f_2 \rceil \circ \beta \circ g_2\Big) = (g_1 \circledast g_2) \circ \langle\langle \pi_1, \pi_2 \circ \pi_2\rangle, \pi_1 \circ \pi_2\rangle \circ (\mathrm{id} \times f_2)$$

$$\Big(\lceil f_1 \rceil \circ \beta \circ g_1\Big) \circledast_f (\alpha \circ g_2) = (g_1 \circledast g_2) \circ \langle \pi_1 \circ \pi_1, \langle \pi_2, \pi_2 \circ \pi_1\rangle\rangle \circ (f_1 \times \mathrm{id})$$

$$\Big(\lceil f_1 \rceil \circ (g_{11} \otimes g_{12})\Big) \circledast_f \Big(\lceil f_2 \rceil \circ (g_{21} \oplus g_{22})\Big) = [g_{11} \circledast g_{21}, g_{12} \circledast g_{22}] \circ \delta_1 \circ (f_1 \times f_2)$$

**Figure 3.** Syntactic definition of the closed-loop composition of systems.

- $s_0 \in \mathbb{S}(T_0, Q, \rho)$ is a state set for the initial parameter type, as given in Definition 5.2; and
- $\tau$ is a set of transitions on $Q$, as described in Definitions 5.4 and 5.5.

An automaton whose state and transition sets are subsets of those of another automaton, with equivalent $\rho$, is called a *subautomaton* or an *automaton fragment*.

**Definition 5.2** (State Sets for Types). Let $T \in \mathbb{T}$ be a type, $Q$ be a finite set of states, and $\rho : Q \longrightarrow \mathbb{T}_\bullet^*$ be a function which provides a list of opaque types for each state in $Q$. A *state set* for $T$ is a function $s : \mathrm{var}\, T \longrightarrow Q$ such that $\forall v \in \mathrm{var}\, T.\ \mathrm{typ}\, v = \rho(s(v))$. The set of all such state sets is denoted $\mathbb{S}(T, Q, \rho)$.

**Definition 5.3** (Variable Mapping). Let $T, U \in \mathbb{T}_\bullet^*$ be lists of opaque types. A *variable mapping* from $T$ to $U$ is a function $m : \mathbb{N}_{|U|} \longrightarrow \mathbb{N}_{|T|}$ assigning an index in $T$ to each index in $U$ such that $\forall i \in \mathbb{N}_{|U|}.\ U_i = T_{m(i)}$. Such a mapping indicates how each variable in a set should be populated from the values of another set of variables, while respecting typing. The set of all such mappings is denoted $\mathbb{M}(T, U)$.

**Definition 5.4** (Computational Transition Group). A *computational transition group* on a state space $Q$ is a 5-tuple $(q_0, f, v_0, m_0, \eta)$, where:

- $q_0 \in Q$ is the source state of the transition;
- $f \in \mathbb{F}_\bullet$ is an opaque function;
- $v_0 \in \mathrm{var}\,(\mathrm{dom}\, f)$ is a variant in the domain of $f$;
- $m_0 \in \mathbb{M}\big(\rho(q_0), \mathrm{typ}\, v_0\big)$ is a variable mapping from the state variables of $q_0$ to $v_0$; and
- $\eta$ is a function which, for each $v_1 \in \mathrm{var}\,(\mathrm{cod}\, f)$, gives a pair $(q_1, m_1)$, where $q_1 \in Q$ is the destination state and $m_1 \in \mathbb{M}\big(\mathrm{typ}\, v_1 \cdot \rho(q_0), \rho(q_1)\big)$, where $\cdot$ denotes concatenation of lists, is a variable mapping from the state variables of $q_0$ and the components of $v_1$ to the state variables of $q_1$.

**Definition 5.5** ($\varepsilon$-Transition). An $\varepsilon$-*transition* on a state space $Q$ is a triple $(q_0, q_1, m)$ of a source state $q_0 \in Q$, a target state $q_1 \in Q$, and a variable mapping $m \in \mathbb{M}\big(\rho(q_0), \rho(q_1)\big)$ from the state variables of $q_0$ to the state variables of $q_1$.

The variable mappings in transitions are generally given in a more natural way by assigning a name to each value which is in scope, as shown in Example 5.6.

The operational semantics of the automaton are as follows. The state of the automaton is a pair $(q, v)$, where $q \in Q$ identifies the state and $x$ is an assignment of correctly-typed values to state variable of $q$, as given by $\rho(q)$. Each outgoing transition represents a valid next step in the computation. Executing a computational transition group means applying the opaque function to the composite value constructed from the source state variables according to the variable mapping, and the resulting state is similarly constructed from the result of the function. Executing an $\varepsilon$-transition simply moves to the target state, constructing its variables according to the variable mapping, without performing any computation (i.e., applying an opaque function to any value).

**Example 5.6.** Suppose we have the state $\alpha$, which is parameterized by types $A$ and $B$; we would write this state as $\alpha(A, B)$. Suppose we similarly have states $\beta(A, C, D)$ and $\gamma(F)$, and an opaque function $\mathsf{foo} : A \times B \longrightarrow C \times D + E \times F$. We can then create a computational transition group on this structure for applying $\mathsf{foo}$ to the parameters of state $\alpha$:

$$\alpha(x, y) \mapsto \begin{cases} \beta(x, w, z) & \kappa_1(w, z) = \mathsf{foo}\,(x, y) \\ \gamma(z) & \kappa_2(w, z) = \mathsf{foo}\,(x, y) \end{cases}$$

It can be easily verified that the expressions in the computation are correctly typed.

*Remark.* In Example 5.6, the component $w : E$ in the second variant of the return type of $\mathsf{foo}$ is discarded. This is perfectly legal, as no part of the model precludes discarding or ignoring values.

**Definition 5.7** (Deterministic and Pseudo-Deterministic Automata). An automaton is called *deterministic* if each state has at most one outgoing $\varepsilon$-transition group, and *pseudo-deterministic* if each state has at most either one outgoing $\varepsilon$-transition or one outgoing computational transition group.

### 5.1 Parallel State

To facilitate the generation of automata, we define the notion of augmenting an automaton fragment with additional state *in parallel*.

**Definition 5.8** (Automaton with Parallel State). Let $A = (Q, \rho, T_0, s_0, \tau)$ be an automaton (or an automaton fragment) and $T \in \mathbb{T}$ be a type; then, $A$ with *parallel state* of type $T$, written $A \parallel_\mathbb{T} T$, is the automaton $(Q \times \text{var}\, T, \rho', T_0 \times T, s'_0, \tau')$, where $\rho'$, $s'_0$ and $\tau'$ are defined as follows:

- $\rho'(q, v) = \rho(q) \cdot \text{typ}\, v$;
- $s'_0(v') = (s_0(v_0), v)$, where $v_0$ and $v$ are the variants of $T_0$ and $T$, respectively, which correspond to $v'$ under the natural isomorphism $\text{var}\,(T_0 \times T) \cong \text{var}\, T_0 \times \text{var}\, T$.
- $\tau'$ is the set which contains, for each $(t, v) \in \tau \times \text{var}\, T$, a transition equivalent to $t$ in which each state $q \in Q$ is replaced by $(q, v)$ and variable mappings are adjusted such that the additional components in each state are simply forwarded along to the corresponding components in the next state.

**Example 5.9.** Suppose we have an automaton fragment with states $\alpha(A)$ and $\beta(B)$, and a single transition:

$$\alpha(x) \mapsto \beta(y),\; y = \text{bar}\,(x)$$

Adding parallel state of type $C \times D + E$ results in an automaton fragment with the states $\alpha_0(A, C, D)$, $\alpha_1(A, E)$, $\beta_0(B, C, D)$, and $\beta_1(B, E)$, and the transitions:

$$\alpha_0(x, y, z) \mapsto \beta_0(w, y, z), \qquad w = \text{bar}\,(x)$$
$$\alpha_1(x, y) \mapsto \beta_1(z, y), \qquad z = \text{bar}\,(x)$$

### 5.2 Implementing Closed-Loop Systems with Automata

This section describes the algorithm by which automata are constructed from closed-loop system specifications.

Observe that, as a result of the syntactic and typing rules given in Section 4.2, all closed-loop system expressions have the form:

$$\lceil \langle \text{func} \rangle \rceil \left( \mu \left( \lceil \langle \text{func} \rangle \rceil \circ \omega \right) \right)$$

Therefore, such systems are fully defined by four values: the initial parameter type, loop state type, initialization function, and loop function. The domain of the initialization function is the initial parameter type, and the domain of the loop function (and codomain of both functions) is the loop state type. The computational behaviour of the system is to apply the initialization function to the initial parameter, then thread the result through infinite applications of the loop function.

Let $T, U \in \mathbb{T}$ be types, $f : T \longrightarrow U$ be a combinational expression, $Q$ be a finite set of existing states, $\rho : Q \longrightarrow \mathbb{T}^*_\bullet$ be a function which provides a list of opaque types for each state in $Q$, $s_0 \in \mathbb{S}(T, Q, \rho)$ be a state set for the domain of $f$, and $s_1 \in \mathbb{S}(U, Q, \rho)$ be a state set for the codomain of $f$. We find $Q'$, $\rho'$ and $\tau'$ as follows, such that $(Q', \rho', T, s_0, \tau')$ is an automaton fragment which implements $f$:

- If $f$ is an opaque function, create a computational transition group from each input state to the corresponding output states for the invocation of the function.
- If $f = \text{id}$, create an identity $\varepsilon$-transition from each input state to the corresponding output state. Note that the input and output types are the same and therefore the state sets will be isomorphic.
- If $f = \theta$, create an $\varepsilon$-transition from each input state to the output state. Note that the output type is 1 and therefore there is exactly one output state, which has no parameters.
- If $f = \pi_1$ (resp. $\pi_2$), create an $\varepsilon$-transition from each input state to the corresponding output state, mapping the corresponding parameters for the left (resp. right) component of the product.
- If $f = \kappa_1$ (resp. $\kappa_2$), create an $\varepsilon$-transition from each input state to the corresponding left (resp. right) output state, leaving no transitions to the right (resp. left) output states.
- If $f = \delta_1$, create an identity $\varepsilon$-transition from each input state to the corresponding output state. Note that, by definition, the input and output types of $\delta_1$ have the same sum-of-products form.
- If $f = f_1 \circ f_2$, create a new state for each variant of the intermediate type (i.e., the type which is both the output of $f_2$ and the input of $f_1$), and recurse into each function accordingly.
- If $f = \langle f_1, f_2 \rangle$, create a new state for the input and output of each function and recurse. Then, add the input to $f_1$ as a parallel state to the resulting structure for $f_2$, and similarly add the output of $f_2$ as a parallel state to the resulting structure for $f_1$. Finally, add $\varepsilon$-transitions to connect: the input states to the input states for $f_2$, duplicating the data into the parallel state parameters; the output states for $f_2$ to the input states for $f_1$, using the actual output parameters as the new parallel state parameters and vice versa; and the output states for $f_1$ to the final output states, matching the output from each function to the corresponding components of the product.
- If $f = [f_1, f_2]$, recurse into each function, passing the same output states, and the appropriate subset of the input states for the left and right of the coproduct.

The procedure described above is used by first creating a state for each variant of the initial parameter type—these are the initial states of the automaton—and for each variant of the loop state type; we will call the latter the *system states*. We then apply the elaboration algorithm to the initialization function, passing the initial states as the input states and the system states as the output states. Finally, we apply the elaboration algorithm to the loop function, passing the system states as both the input states and output states. The result is a pseudo-deterministic automaton which represents the exact computation performed by the closed-loop system.

### 5.3 ε-Elimination

Similarly to the well-known algorithm for nondeterministic finite automata [15, p. 26], we can eliminate $\varepsilon$-transitions from an automaton by recursively replacing them with all outgoing transitions from their target state. For our construction, this process must include mapping the correct parameters, effectively composing the (trivial) function represented by the $\varepsilon$-transition to the right of any transitions which it is being replaced with. Here, we present an algorithm which does this for pseudo-deterministic automata.

**Definition 5.10** ($\varepsilon$-Elimination in Pseudo-Deterministic Automata). Let $A = (Q, \rho, T, s_0, \tau)$ be a pseudo-deterministic automaton and $r_\tau : Q \twoheadrightarrow \tau$ be the partial function which gives the unique outgoing transition from a state, if it exists. Performing $\varepsilon$-elimination on $A$ results in an automaton $\hat{A} = \left(Q, \rho, T, s_0, \{R(q, \emptyset, \text{id}, r_\tau(q)) \mid q \in Q\}\right)$, where the partial function $R$ is defined as follows:

$$R\left(q_0, s, \hat{m}, (q'_0, f, v_0, m_0, \eta)\right) = (q_0, f, v_0, \hat{m} \circ m_0, \hat{\eta}),$$

$$\text{where } \hat{\eta}(v_1) = (q_1, \hat{m}_1 \circ m_1),$$

$$(q_1, m_q) = \eta(v_1),$$

$$\hat{m}_1(i) = \begin{cases} i & i < n \\ \hat{m}(i-n) + n & i \geq n \end{cases},$$

$$n = |\text{typ } v_1|$$

$$R\left(q_0, s, \hat{m}, (q'_0, q_1, m)\right) = R\left(q_0, s \cup \{q'_0\}, \hat{m} \circ m, r_\tau(q_1)\right),$$

$$\text{if } q'_0 \notin s$$

The function $R$ recursively rewrites $\varepsilon$-transitions into computational transition groups by tracing the path of $\varepsilon$-transitions through the automaton (using $r_\tau$) and composing together the variable mappings in order to properly generate the mappings in the resulting transition. States which have no outgoing transitions, or have an outgoing sequence of $\varepsilon$-transitions which do not lead to a computational transition group, are left with no outgoing transitions in $\hat{A}$.

**Lemma 5.11.** *In Definition 5.10, the recursive function $R$ always terminates.*

*Proof.* As the sole recursive call passes $s \cup \{q'_0\}$ for $s$ and is guarded by $q'_0 \notin s$, the cardinality of $s$ increases by 1 each recursive call. Since $s \subseteq Q$ and $Q$ is finite, the maximum depth of recursion is equal to $|Q|$ and thus finite, and therefore the algorithm terminates. □

### 5.4 Similarity

Similarity of automata is formulated such that an automaton $A_1$ is similar to an automaton $A_2$ if there is a correspondence from states and state variables in $A_1$ to states and state variables in $A_2$, such that all computations which may be performed at a state in $A_1$ may also be performed on the corresponding state variables at a corresponding state in $A_2$. It is formally defined as follows:

**Definition 5.12** (Similarity of Automata). Given two automata $A_1 = (Q_1, \rho_1, T, s_{0,1}, \tau_1)$ and $A_2 = (Q_2, \rho_2, T, s_{0,2}, \tau_2)$ with the same initial parameter type $T$ and no $\varepsilon$-transitions, a *simulation* of $A_1$ in $A_2$ is a relation $R \subseteq \left(Q_1 \times \mathbb{N}^{\mathbb{N}}\right) \times \left(Q_2 \times \mathbb{N}^{\mathbb{N}}\right)$ between pairs $(q_1, m_1)$ and $(q_2, m_2)$, which obeys the following axioms:

1. $\forall v \in \text{var } T. \left((s_{0,1}(v), \text{id}), (s_{0,2}(v), \text{id})\right) \in R$
2. For all $\left((q_1, m_1), (q_2, m_2)\right) \in R$ and $(q_1, f, v_0, m_0, \eta) \in \tau_1$, there exists in $\tau_2$ a computational transition group $\left(q_2, f, v_0, m'_0, \eta'\right)$ such that $m_1 \circ m_0 = m_2 \circ m'_0$ and for each $v_1 \in \text{var}(\text{cod } f)$, where $\left(q'_1, m'_1\right) = \eta(v_1)$ and $\left(q'_2, m'_2\right) = \eta'(v_1)$, there exists in $R$ an element $\left(\left(q'_1, m''_1\right), \left(q'_2, m''_2\right)\right)$, where for any indices $i$ and $j$, $m''_1(i) = m''_2(j)$ if and only if $m'_1(i)$ and $m'_2(j)$ refer to either the same component in $v_1$ or state variables $i'$ and $j'$ (in $q_1$ and $q_2$) such that $m_1(i') = m_2(j')$.

The simulation thus describes how states and state variables in $A_1$ correspond to states and state variables in $A_2$. Axiom 1 requires that respective initial states correspond to each other; axiom 2 requires that computational transition groups in $A_1$ are reflected in $A_2$.

If a simulation exists, then $A_1$ is *similar* to $A_2$, denoted by infix $\lesssim$. Similarity is extended to automata with $\varepsilon$-transitions by noting that similarity is transitive and defining that, for any automaton $A$, $A \lesssim \hat{A}$ and $\hat{A} \lesssim A$.

**Lemma 5.13.** *Similarity of automata is decidable.*

*Proof.* Let $A_1$ and $A_2$ be automata with the same initial parameter type $T$, and suppose $R$ is a simulation of $A_1$ in $A_2$. We may define an equivalence relation $\equiv$ on $R$ such that $\left((q_1, m_1), (q_2, m_2)\right) \equiv \left((q_1, m'_1), (q_2, m'_2)\right)$ if and only if, for all indices $i$ and $j$, $m_1(i) = m_2(j) \Leftrightarrow m'_1(i) = m'_2(j)$, i.e., they represent the same correspondence between state variables for the same pair of states. It can be seen that elements of the same equivalence class under $\equiv$ are treated the same by the simulation axioms, and that there are finitely many equivalence classes. Thus, the existence of $R$ may be checked by attempting to recursively enumerate $R/\equiv$ using the axioms: if at any point, the transitions required by axiom 2 are not present in $A_2$, then $A_1 \not\lesssim A_2$; otherwise, if the enumeration process terminates successfully, then $A_1 \lesssim A_2$. □

### 5.5 Commutative Extension

In this section, we develop a transformation which will prove useful for the verification of closed-loop systems. Given a sequence of transitions in an automaton, we wish to allow the computations represented by later transitions to occur earlier if the values which they operate on are available (i.e., do not depend on the results of the intervening computations). An algorithm is given for computing the commutative extension of a deterministic automaton.

**Definition 5.14** (Commutative Extension of an Automaton). For any automaton $A = (Q, \rho, T, s_0, \tau)$ and value $x_0 : T$, let $\Diamond_\mathcal{A}(A, x_0)$ be the set of all pairs $(f, x)$, where $f \in \mathbb{F}_\bullet$ and $x : \text{dom } f$, such that $A$ will eventually apply the function $f$ to the value $x$. The *commutative extension* of $A$, denoted $\tilde{A}$, is the largest automaton (with respect to similarity) such that $\forall x_0 : T, \Diamond_\mathcal{A}(\tilde{A}, x_0) = \Diamond_\mathcal{A}(A, x_0)$; if $A$ is deterministic, then $\tilde{A}$ is computed as follows.

Let $r_\tau$ be as given in Definition 5.10, and for any state $q \in Q$, let $\square(q)$ be the smallest set of 3-tuples such that:

$$\square(q) = \{(f, v_0, m_0, \varepsilon) \mid (q, f, v_0, m_0, \eta) \in \tau\} \cup$$
$$\{(f', v'_0, m'_0, L) \mid (q, f, v_0, m_0, \eta) \in \tau \wedge P\},$$

where $P \Leftrightarrow (q_1, m_1) = \eta(v_1) \wedge \forall v_1 \in \text{var}(\text{cod } f)$.

$$L(v_1) \in \square(q_1) \wedge m_1 \circ m = \lambda i.\, m'_0(i) + |\text{typ } v_1| \wedge$$
$$(f', v'_0, m, L') = L(v_1)$$

Here, $\square(q)$ describes the set of all computations which are guaranteed to be performed by the automaton after it is in state $q$, and which depend only on the variables of that state; thus, these are the computations which may be performed by the commutative extension in (the image of) state $q$ without changing the behaviour. By construction, $\square(q)$ is finite for all $q \in Q$, and therefore it may be computed incrementally from $\tau$ using the above definition.

Using this function, $\tilde{A}$ is constructed as follows. We replace each state $q \in Q$ with an acyclic subautomaton which has a state for each pair of a subset $E \subseteq \square(q)$ and a mapping $r$ from entries $(f, v_0, m_0, L) \in E$ to a codomain variant $v_1 \in \text{var}(\text{cod } f)$ of the corresponding opaque function. These states correspond to the possible progress and outcomes of performing the computations in $\square(q)$; the state with $E = \emptyset$ corresponds to the same computational progress as $q$ in the original automaton, and each entry in $E$ indicates an additional computation which has been performed. Accordingly, each state in the subautomaton has the list of state variables $\rho(q) \cdot \text{typ } r(e_1) \cdot \text{typ } r(e_2) \cdots \text{typ } r(e_n)$, where $E = \{e_1, e_2, \ldots e_n\}$, and an outgoing computational transition group for each $e \in \square(q) \setminus E$ with the destination states being those $(E', r')$ with $E' = E \cup \{e\}$ and $r'$ extending $r$ with the result of the computation. The transitions have the necessary variable mappings to retain the values of the original state variables and of computation results in corresponding locations.

Observe that any $(f, v_0, m_0, L) \in \square(q)$ is either the unique entry corresponding to the outgoing transition from $q$ in $A$ (with the special value $L = \varepsilon$) or is inherited as an entry common to all the destination states of that transition. Using this information, the sub-automata can be connected as follows to finish constructing $\tilde{A}$.

Let $(q, f, v_0, m_0, \eta) = r_\tau(q)$, $\overline{e} = (f, v_0, m_0, \varepsilon)$ be the unique such entry in $\square(q)$, and $(q_1, m_1) = \eta(r(\overline{e}))$. From each sub-automaton state $(E, r)$ where $\overline{e} \in E$, we create an outgoing $\varepsilon$-transition to the state $(E', r')$ in the subautomaton for $q_1$ where $E' = \{L(e) \mid e \in E \setminus \{\overline{e}\}\}$ and $r'(L(e)) = r(e)$ for all $e \in E \setminus \{\overline{e}\}$. Similarly to the computational transition groups, each $\varepsilon$-transition has the necessary variable mapping to propagate computation results (for all $e \in E'$) and to populate the state variables of $q'$ from those of $q$ and the results of $\overline{e}$.

The commutative extension enables the following result, which is crucial to Theorem 5.16.

**Lemma 5.15.** *Let $T \in \mathbb{T}$ be a type, $S_1$ and $S_2$ be closed-loop systems both having initial parameter type $T$, and $A_1$ and $A_2$ be the respective automata implementing these systems. Then, $S_1 \lesssim_\Diamond S_2$ if and only if $A_1 \lesssim \tilde{A}_2$.*

*Proof.* From Definitions 4.3 and 5.14, it is clear that, for all $x_0 : T$, $\Diamond_\mathcal{A}(A_i, x_0) = \Diamond(S_i, x_0)$ for each system. Thus, the proof reduces to showing that $\Diamond_\mathcal{A}(A_1, x_0) \subseteq \Diamond_\mathcal{A}(A_2, x_0)$ if and only if $A_1 \lesssim \tilde{A}_2$; the proof sketch of this is as follows.

(Only if). Let $(f, x) \in \Diamond_\mathcal{A}(A_1, x_0)$, i.e., $A_1$ (when initialized with $x_0$) applies $f$ to $x$ at some point—and by assumption, so does $A_2$. Then each automaton has a state which at some point contains the components of $x$ in its state variables, and which has an outgoing computational transition group which performs the computation. Since each component in $x$ is either a component of $x_0$ or is computed from $x_0$ using finite applications of opaque functions, the simulation axioms are satisfiable by induction on how the components of $x$ were computed.

(If). Let $(f, x) \in \Diamond_\mathcal{A}(A_1, x_0)$. Then $A_1$ has a state which at some point contains the components of $x$ in its state variables, and which has an outgoing computational transition group which performs the computation. By Definition 5.12, such a state also exists in $\tilde{A}_2$, and $\Diamond_\mathcal{A}(\tilde{A}_2, x_0) = \Diamond_\mathcal{A}(A_2, x_0)$; therefore, $(f, x) \in \Diamond_\mathcal{A}(A_2, x_0)$. □

**Theorem 5.16.** *Similarity of closed-loop systems is decidable.*

*Proof.* Immediate from Lemmas 5.13 and 5.15. □

## 6 Modelling the Outside World

Many real-world applications of synchronous digital logic design involve components which interface with external components, such as storage devices, computer networks, or humans. Incorporating external components into our model has two main challenges: firstly, the precise behaviour of these systems is often complex (especially in the case of a human), and secondly, verification has only been defined thus far for closed-loop systems, which supposedly have no inputs or outputs.

The solution to these problems is to note that—from the perspective of a digital control system—the entire outside world which exists behind an interface may well be contained inside that interface; thus, we model the external components as a value of an opaque type stored inside the system, representing the "state of the world". Any interactions with the external components are modelled as opaque functions, which take as one of their inputs the current state of the world and produce as an output a new state of the world. The system must then be designed such that the old value is discarded after being used, as there is no way to access the "world before the interaction occurred" after it occurs. Different patterns of behaviour on the part of the external component—such as a human pressing a key, or not—can be represented by variants in the codomain of the opaque function.

**Example 6.1.** Suppose we have a communication interface which permits us to either send a message or attempt to receive a message each cycle, at our choosing. We wish to design a controller which attempts to receive a message every cycle, unless a message was received in the previous cycle, in which case that same message is sent.

Suppose $\mathrm{MSG}, \mathrm{WRLD} \in \mathbb{T}_\bullet$, where MSG is the type of messages and WRLD is the type which represents the outside world, and there exist opaque functions $\mathrm{send} : \mathrm{WRLD} \times \mathrm{MSG} \longrightarrow \mathrm{WRLD}$, which sends a message, and $\mathrm{recv} : \mathrm{WRLD} \longrightarrow \mathrm{WRLD} \times (\mathrm{MSG} + 1)$, which attempts to receive a message. The specification of our desired behaviour is given as:

$$S = \lceil \mathrm{id} \rceil \left( \mu \left( \lceil [\mathrm{send}, \pi_1] \circ \delta_1 \circ \mathrm{recv} \rceil \circ \omega \right) \right)$$

The communication interface (i.e., the plant) has the shape $\mathrm{Id}^{\mathrm{MSG}} \times \mathrm{Id}_{\mathrm{MSG}+1}$; the product functor permits the selection of sending or receiving mode each cycle. The plant is described by the following system expression:

$$P = \lceil \mathrm{id} \rceil \left( \mu \left( (\alpha \circ \lceil \mathrm{send} \rceil \circ \omega) \otimes (\lceil \mathrm{recv} \rceil \circ \beta \circ \omega) \right) \right)$$

The controller must then have the complementary shape $\mathrm{Id}_{\mathrm{MSG}} + \mathrm{Id}^{\mathrm{MSG}+1}$; we wish to design a controller $C$ such that $S \lesssim_\diamond P \circledast C$. The following controller can be shown to satisfy this specification:

$$C = \lceil \kappa_2 \circ \theta \rceil \left( \mu \left( \left( \lceil \langle \kappa_2 \circ \theta, \mathrm{id} \rangle \rceil \circ \beta \circ \omega \right) \oplus \left( \alpha \circ \lceil \pi_2 \rceil \circ \omega \right) \right) \right)$$

This approach to representing interactions with the outside world within a model of pure computations is an exact analogue of that which was introduced to functional programming by Launchbury and Peyton Jones [19]. In that case, the proper threading of an opaque value representing the outside world through functions which perform impure operations is ensured by the IO monad, whereas here it is enforced by the system specification for the plant.

While this approach does place some responsibility on the author of the plant specification to properly implement this pattern, it is more flexible than the monadic version as a result. For example, non-destructive interactions with the outside world—those which do not invalidate the state of the world, such as the example given in the introduction of reading from memory—may be modelled by the plant remembering the prior state, so that the timing-agnostic nature of consecutive non-destructive operations is recognized by the semantics of the model when checking similarity.

## 7 Future Work

While the models presented in this paper are sufficient for verifying solutions to a significant class of control problems, there are a few directions in which further work would result in greater utility.

Firstly, the type system and computational model could be extended with support for definitional equalities and similar logical extensions. This would enable the model to be informed that, for example, the arguments to a commutative function such as addition may be swapped without affecting the result, or that a read immediately following a write at the same address will produce the value which was written.

Secondly, in both the system and automaton models, the support of some form of modularity of which the decision algorithm is aware has the potential to significantly improve the efficiency of that algorithm; as our model, unlike many in functional programming, intentionally omits first-class functions in order to ensure finiteness of implementations, we do not get a form of modularity for "free", and must instead introduce it explicitly. Modularity could in principle be further extended to forms of polymorphism, as has been done in discrete-event systems using templates [13] and in functional programming with parametric polymorphism [12].

There is also potential to develop additional means of composing systems together in order to describe a wider variety of architectures. This paper only contemplates connecting a single controller to a single plant; new composition operators could enable the combination of multiple sub-plants to form a larger plant, or consider situations in which there are multiple controllers with different capabilities, such as with the decentralized control of discrete-event systems [23].

Finally, solving the control problem—possibly under certain constraints—would turn this model into a very powerful tool for the design of digital systems. The control problem may in turn be extended to a variety of optimal control problems, as has been done for discrete-event systems [27].

## Acknowledgments

We acknowledge the support of the Natural Sciences and Engineering Research Council of Canada (NSERC), funding reference numbers USRA–563528–2021 and RGPIN–2020–04279.

Cette recherche a été financée par le Conseil de recherches en sciences naturelles et en génie du Canada (CRSNG), numéros de référence USRA–563528–2021 et RGPIN–2020–04279.